\documentclass[pra,twocolumn,amsmath,superscriptaddress,aps,floatfix,showpacs]{revtex4}

\usepackage{graphicx}
\usepackage{color}
\usepackage{marvosym}
\usepackage{wasysym}
\usepackage{pifont}
\usepackage{amssymb}
\usepackage{hyperref}

\newcommand{\comment}[1]{}
\newcommand{\fys}{Department of Engineering Physics, Helsinki University of
  Technology, P.O.~Box 5100,
FI-02015 TKK, Finland}
\newcommand{\cqct}{Australian Research Council Centre of Excellence for Quantum Computer Technology,
The University of New South Wales, Sydney 2052, Australia}
\newcommand{\LTL}{Low Temperature Laboratory, Helsinki University of Technology,
  P.O.~Box 3500,
FI-02015 TKK, Finland}
\newcommand{\bristol}{Centre for Quantum Photonics, H. H. Wills Physics Laboratory \& Department of Electrical and Electronic Engineering,
University of Bristol, Merchant Venturers Building, Woodland Road,
Bristol, BS8 1UB, UK}

\begin{document}

\title{Entanglement-Enhanced Quantum Key Distribution}

\author{Olli Ahonen}\affiliation{\fys}
\author{Mikko M\"ott\"onen}\affiliation{\fys}\affiliation{\cqct}\affiliation{\LTL}
\author{Jeremy O'Brien}\affiliation{\bristol}
\date{\today}

\begin{abstract}
We present and analyze a quantum key distribution protocol based
on sending entangled $N$-qubit states instead of single-qubit ones
as in the trail-blazing scheme by Bennett and Brassard (BB84).
Since the qubits are sent and acknowledged individually, an eavesdropper is limited
to accessing them one by one. In an intercept-resend attack, this
fundamental restriction allows one to make the eavesdropper's
information on the transmitted key vanish if even one of the
qubits is not intercepted. The implied upper bound $1/(2N)$ for
Eve's information is further shown not to be the lowest, as the
information can be reduced to less than 30\%
of that in BB84 in
the case $N=2$. In general, the protocol is at least as secure as BB84.
\end{abstract}

\hspace{5mm}


\maketitle

\section{Introduction}

Quantum information science \cite{nielsen} has emerged to answer
the question: ``What additional power and functionality can be
gained by processing and transmitting information encoded in
physical systems that exhibit uniquely quantum mechanical
behavior?" Anticipated future quantum technologies include:
quantum computing \cite{nielsen,de-prsla-400-97}, which promises
exponential speed-up for particular computational tasks; quantum
metrology \cite{gi-sci-306-1330}, which allows the fundamental
precision limit to be reached; and quantum lithography
\cite{bo-prl-85-2733}, which could enable fabrication of devices
with features much smaller than the wavelength of light.
 The most striking quantum technologies that have already reached commercial realization are in the area of quantum communication.

Quantum key distribution (QKD) offers secure communication based
on the fundamental laws of physics---namely, that measurement of a
quantum system being used to transmit information must necessarily
disturb that system, and that this disturbance is detectable
\cite{gi-rmp-74-145}. The first QKD scheme was proposed by Bennett
and Brassard in 1984 (BB84) and is based on generating a cryptographic secret
key between two distant parties, Alice and Bob, by sending a random bit string encoded
and measured in one of two randomly chosen mutually unbiased bases
of a single qubit \cite{bb84}. Photons are the logical choice for
transmitting quantum information and were used in the first
experimental realization of BB84 \cite{be-jcrypt-5-3}. Since then
there have been several important theoretical improvements and
experimental demonstrations of BB84 and other QKD protocols
\cite{e91, mod94, mod95, mod98, mod05a, mod05b, mod05c, mod06a,
mod06b}, which have culminated in commercial QKD systems. A major
challenge facing future practical quantum networks is to increase
the rate at which the secure key is generated. Most efforts in
this direction are focused on improving the underpinning
technology \cite{gi-rmp-74-145}. Here we propose an alternative
approach based on improving the underlying QKD protocol, which has
been inspired by recent developments in optical quantum computing
\cite{ob-sci-318}.

The ability to reliably entangle photons is a major goal of
quantum information processing \cite{ob-sci-318} and quantum
communication. Recent demonstrations of strong coupling between
semiconductor quantum dots and photonic crystal cavities has been
reported \cite{yo-nat-432-200,he-nat-445-896,en-nat-450-857}. The
generation and transfer of photons on a photonic crystal chip has
been demonstrated \cite{en-oe-15-5550}, together with entangling
photonic logic gates all in fibre \cite{arxiv-08021676}, and in
waveguides on silicon chips \cite{po-sci-320-646}. The breakthrough
proposal based on measurement induced nonlinearities
\cite{kn-nat-409-46}, capable of entangling photons for optical
quantum computing, was followed by important demonstrations of
entangling logic gates \cite{ob-nat-426-264, ob-prl-93-080502,ga-prl-93-020504}.
Recently, attention has
focused on generating entangled
states of many photons, and it was shown that atom-cavity
systems can be used to generate an arbitrary entangled state of
$N$ photons \cite{de-pra-76-052312}. Thus the technology for performing
an entangling transformation on several photons is now within sight.

Here we present a novel QKD protocol whose security is
lower-bounded by BB84. The insight of the procotol relies on Alice entangling groups of
qubits prior to their one-by-one transmission. Because successive
qubits in each group are transmitted only after confirmation of reception by
Bob, an eavesdropper only has access to the transmitted
information one qubit at a time. The eavesdropper is thus unable
to perfectly undo the entangling transformation even if aware of
it. Qubits from different entangled groups can be sent interleaved to
keep the quantum channel utilization high. We present the
maximal mutual information on the established
key provided by any intercept-resend (IR) attack, and also the
corresponding induced disturbance, quantified by the quantum bit
error rate (QBER), for several entangling transformations. We show
that only small groups of qubits need to be entangled for
substantial gains: Utilizing two-qubit entanglement, it is
possible to significantly reduce an eavesdropper's maximal
information on the key, e.g., to less than 30\% of that in BB84
for a fixed $\mathrm{QBER} \leq 25\%$. Furthermore, another
multi-qubit entangling transformation reduces the information gain
to zero in the case where the IR attacker intercepts all but one
of the qubits, which is shown to restrict the maximal information
gain to $1/(2N)$. Finally, we present a rough estimate of the key
generation rate for the optimal two-qubit protocol.

\section{The Protocol}

In our protocol, the initiator, Alice, generates a number of
random bits, handled in groups of $N$. Each group is an outcome of
the random variable $A = A_1A_2\cdots{}A_N$ composed of the binary
random variables $A_i$, for which the probabilities are $p(A_i =
0) = p(A_i = 1) = \frac{1}{2}$, $i = 1,\ldots,N$. Let the bit
string $a = a_1a_2\cdots{}a_N$ denote the outcome of $A$. These
bits form Alice's raw key.

Alice uses a public quantum channel to transmit the raw key to the
recipient, Bob. The basis of each qubit is random, being the
eigenbasis of the Pauli matrix $\sigma_z$, $\{|0\rangle,
|1\rangle\}$ or that of $\sigma_x$, $\{|+\rangle = (|0\rangle +
|1\rangle)/\sqrt{2}, |-\rangle = (|0\rangle -
|1\rangle)/\sqrt{2}\}$ with equal probability. Let $\alpha =
\alpha_1\alpha_2\cdots\alpha_N$, with each $\alpha_i \in \{z,x\}$,
denote Alice's basis choices for an $N$-bit group.

\begin{figure}[h]
\centering
\includegraphics[width=0.48\textwidth]{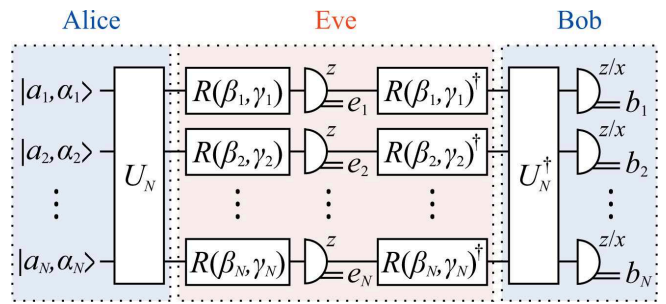}
\caption{\label{circuit} Quantum circuit for the proposed protocol
with the intercept-resend attack. The circuit is repeated until a
large enough number of bits has been transmitted. The required
classical communication is not shown. Semicircles represent
projective measurements. Gate $U_N$ is an $N$-qubit entangler
announced in public. Gate $R(\beta,\gamma)$ rotates a qubit by the
angles $\gamma$ and $\beta$ with respect to $z$ and $y$ axes,
respectively. Each $a_i$, $e_i$, and $b_i$ is a binary variable
representing a bit in Alice's, Eve's, and Bob's raw key,
respectively.}
\end{figure}

Before transmission, Alice applies a fixed $N$-qubit gate $U_N$,
declared in public, to each group
$\{|a_i;\alpha_i\rangle\}_{i=1}^N$. Thus the qubits are, in
general, entangled. She then sends the qubits one by one to Bob,
always waiting for Bob to acknowledge each qubit on a public
authenticated classical channel before sending the next one. This
waiting does not decrease the transmission rate: Individual qubits
from different groups can be sent interleaved. Bob waits for $N$
qubits to accumulate, and applies $U_N^{\dagger}$ to the group. He
projectively measures each qubit in the $\sigma_z$ or $\sigma_x$
eigenbasis, chosen at random, and obtains his raw key, consisting
of the measurement results $b_i \in \{0,1\}$. Figure \ref{circuit}
shows the protocol as a quantum circuit for the $N$ qubits. The
quantum non-demolition (QND) measurements needed for Bob to detect
the reception of each qubit are not shown. The QND measurements
can be performed with high fidelity, as is demonstrated, for
instance, in Ref.~\cite{pr-prl-92-190402}.

After the quantum transmission, Alice and Bob compare their basis
choices over the classical channel, and discard the raw-key bits
for which their bases did not coincide. Note that the entire
$N$-bit group need not be discarded, only the individual
incompatible results. The remaining bits form the participants'
sifted keys which may still contain differences due to noise or
eavesdropping on the quantum channel. Based on these differences,
Alice and Bob estimate the QBER, defined formally in Sec.~\ref{analysis}. If the
observed QBER is less than 15\%, errors can be corrected by a
classical error correction (EC) procedure, e.g., by one
 described
in Ref.~\cite{cascade}. If eavesdropping is suspected, Alice and
Bob employ privacy amplification which shortens the key and
reduces any eavesdropper's information on it to an arbitrarily low
value. For QBER's in the range 15-25\%, less efficient quantum privacy
amplification or classical advantage distillation techniques can be used to
arrive at a secure and error-free key \cite{gi-rmp-74-145}.

\section{Analysis}
\label{analysis}

First, we point out that our protocol cannot be less secure than
BB84, even if Eve is allowed any attack strategy. Giving Eve
 full control of the gates $U_N$ and
$U_N^{\dagger}$ shown in Fig.~\ref{circuit} reduces the protocol
to BB84 facing a coherent
attack. Thus, the proofs of security for BB84 with coherent attacks allowed
(Ref.~\cite{bb84proof07} and references therein) also apply to our
protocol, and Alice and Bob can ensure the secrecy of the generated
key in our protocol, as well.

\begin{figure}[h]
\centering
\includegraphics[width=0.36\textwidth]{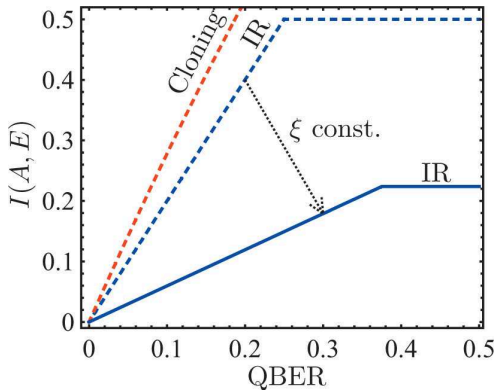}
\caption{\label{infoqber} Eve's information per bit on Alice's
sifted key as a function of the observed QBER for BB84 with
cloning and intercept-resend (IR) attacks (dashed lines), and for
our protocol using $U_2 =
C\!\left(\frac{\pi}{32},\frac{3\pi}{8},\frac{\pi}{32}\right)$ with
the corresponding optimal IR attack (solid line). The arrow shows
the effect of engaging $U_2$ while keeping the fraction of
intercepted qubits $\xi = 0.8$ constant.}
\end{figure}

We continue our more refined analysis by studying the protocol under the IR
attack. Potentially more efficient, e.g., cloning, attacks are to be
studied in future work.
In all attacks, the goal of the attacker is to obtain a copy of
the sifted key for a minimal increase in the QBER, which is the
only indicator of careful eavesdropping to Alice and Bob. In BB84, the IR
attack is succinctly described as the eavesdropper, Eve, measuring
the transmitted qubits in $z$ or $x$ basis and resending the
obtained results to Bob. Independent of Eve's choice of basis, she
obtains on average at most 0.5 bits of information on each bit of
the sifted key, and induces an average QBER of at least 25\%
\cite{bb84}. A slightly better strategy for Eve is to clone each
qubit imperfectly and measure the clone state \cite{cloneattack}.
The more information Eve extracts on the key, the larger the
induced error rate is. Eve can also choose to interfere only with
a fraction $\xi \in [0,1]$ of the transmitted qubits. Eve's maximal
information as a function of QBER is shown in Fig.~\ref{infoqber}
for these attacks.

In our protocol, Eve's choice of basis has a significant impact on
her information and the induced QBER. Hence, we allow Eve to
measure each qubit in any basis. This is equivalent to allowing
Eve arbitrary single-qubit gates, and measurements in the $z$
basis. For the group of $N$ qubits, Eve's measurement results are
the outcomes $e = e_1e_2\cdots{}e_N$ of the random variable $E$,
with each $e_i \in \{0,1\}$.

Once Eve has measured a qubit, the result $e_i$ represents her
best guess on Alice's corresponding key bit. Therefore, to
minimize the QBER, she constructs the state $|e_i;z\rangle$ and
then undoes the previously applied single-qubit gate before
sending the qubit to Bob. Any single-qubit gate can be written as
three successive rotations about the Bloch-sphere axes $y$ and
$z$, $R_z(\varphi) R_y(\beta) R_z(\gamma) e^{i\phi}$. Since Eve
measures in the $z$ basis, the final rotation $R_z(\varphi)$ has
no effect on the result. The global phase $\phi$ is irrelevant as
well. Eve's attack is thus parametrized by the single-qubit gate
rotation angles
$\{(\beta_1,\gamma_1),\ldots,(\beta_N,\gamma_N)\}$.

The information Eve gains on the key is quantified by the mutual
information of the random variables $A$ and $E$, defined
as~\cite{nielsen}
\begin{equation}
I(A,E) = \frac{1}{N}[H(A) + H(E) - H(A,E)],
\label{eq:mutualinformation}
\end{equation}
where $H(\cdot)$ denotes the Shannon entropy
 and $H(\cdot,\cdot)$ the joint entropy. The factor $\frac{1}{N}$ ensures that Eq.~(\ref{eq:mutualinformation}) yields the mutual information per bit, since $A$ and $E$ are both $N$-bit entities. The entropies must be averaged over Alice's choice of bases $\alpha$ which Eve eventually finds out. Thus,
$H(A,E)  = \frac{1}{2^{N}} \sum_{\alpha} H_{\alpha}(A,E)
            = - \frac{1}{2^{N}} \sum_{\alpha,a,e} p(a,e|\alpha) \log_2
            p(a,e|\alpha)$,
and $H(E) = \frac{1}{2^{N}} \sum_{\alpha} H_{\alpha}(E) = -
\frac{1}{2^{N}} \sum_{\alpha,e} p(e|\alpha) \log_2 p(e|\alpha)$,
where the probabilities are conditioned on $\alpha$. The entropy
$H(A) = N$.

The QBER is defined as the average probability of a bit flip in
the sifted key. For each individual qubit $j = 1,\ldots,N$ it is
\begin{equation}
\mathrm{QBER}_j = \frac{1}{4} \sum_{\alpha_j = z}^x \sum_{a_j =
0}^1 p(B_j = \bar{a}_j | A_j = a_j; \alpha_j), \nonumber
\end{equation}
where $B_j$ is the random variable giving Bob's measurement result
$b_j$ of $j$th qubit, and the bar denotes the logical \textsc{not}
operation. The QBER used in the following analysis is the average
of the QBER's of the $N$ qubits.

For Alice and Bob to accept the sifted key for post-processing, the
fraction of eavesdropped qubits $\xi$ must be such that
$\mathrm{QBER} \leq 0.25$. Typically, they set a suitable threshold
value for acceptance \cite{gi-rmp-74-145}
in this regime, where the information gain of
the eavesdropper is linear with respect to QBER in the IR attack.
Therefore, Eve's maximal information for a given QBER is
determined by the maximum of the ratio $I(A,E)/\mathrm{QBER}$.

The final bit rate $R_{\mathrm{net}}$ is an important measure of
efficiency for a QKD protocol. This is the rate at which Alice and
Bob accumulate shared secret key bits, which contain no errors,
and on which Eve's information is negligible, i.e., below a known
bound controlled by Alice and Bob. Since the transformations $U_N$
and $U_N^{\dagger}$ provide no new capabilities for Eve under the
coherent attack model for BB84, the final bit rate of our protocol
cannot be lower than in BB84, with an ideal quantum channel.
However, innocent noise in the quantum channel may change this
setting.

Let us present a recursive construction for the gate $U_N$ which
bounds the information of an IR attacker to at most $1/(2N)$ for
any QBER, a proof of which is given in the Appendix. We denote this
gate by $U_N^{\displaystyle{\star}}$.
The gate has two equivalent
versions of different parity: $U_{N,\mathrm{even}}^{\displaystyle{\star}}$
and $U_{N,\mathrm{odd}}^{\displaystyle{\star}}$, either one can be used as $U_N^{\displaystyle{\star}}$. We define
$U_{1,\mathrm{even}}^{\displaystyle{\star}} = I_1$, the one-qubit identity
operation, and $U_{1,\mathrm{odd}}^{\displaystyle{\star}} = \sigma_y$. The unitary $(N+1)$-qubit gate is obtained with the
following rule:
\begin{equation} U_{N+1}^{\displaystyle{\star}} = \frac{1}{\sqrt{2}}
\left[I_1 \otimes U_N^{\displaystyle{\star}} \pm i \sigma_y \otimes
\left(P_N U_N^{\displaystyle{\star}}\right)\right], \label{ngate}
\end{equation}
where $P_N = \sigma_{y} \otimes I_1^{\otimes N-1}$ if $N\geq2$ and $P_1 = \sigma_y$.
At each step, either of the two signs can be chosen.

The fact that, with gate $U_N^{\displaystyle{\star}}$, Eve cannot
miss even a single qubit unless she is content with zero
information gain also protects the key distribution against
photon-number splitting (PNS) attacks \cite{pnsattack}. If the
probability of an unwanted multi-photon pulse is $\varepsilon$ and
events are independent, the probability that Eve gains any
information decreases at least as $\varepsilon^N$.

In what follows, we study the case $N=2$ in more detail. Arbitrary two-qubit
gates have 16 degrees of freedom, several of which have no effect
on Eve's maximal information. First fixing the global phase of the
gate and then following the treatment in Ref.~\cite{prazhang}, we
obtain $U_2 =\, (k_{2,1} \otimes k_{2,2})
           \times \exp \left[ \frac{i}{2} \left(c_1\, \sigma_x \otimes \sigma_x + c_2\, \sigma_y \otimes \sigma_y + c_3\, \sigma_z \otimes \sigma_z\right) \right]
            \times (k_{1,1} \otimes k_{1,2})$,
where $k_{j,l}$ are one-qubit gates and the middle gate,
$C(\mathbf{c})$, has parameters $\mathbf{c} = (c_1,c_2,c_3)$ with
each $c_j \in [0,2\pi]$. The local operation $k_{2,1} \otimes
k_{2,2}$ can be directly undone by Eve, and is thus of no use to
Alice and Bob. Hence, the interesting two-qubit gates are of the
form $C(\mathbf{c})(k_{1,1} \otimes k_{1,2})$. To simplify the
calculations, we set $k_{1,1} = k_{1,2}$.
Removing this restriction can only improve the results presented
in Sec.~\ref{results}.

\section{Results}
\label{results}

\begin{figure}[h]
\centering
\includegraphics[width=0.4\textwidth]{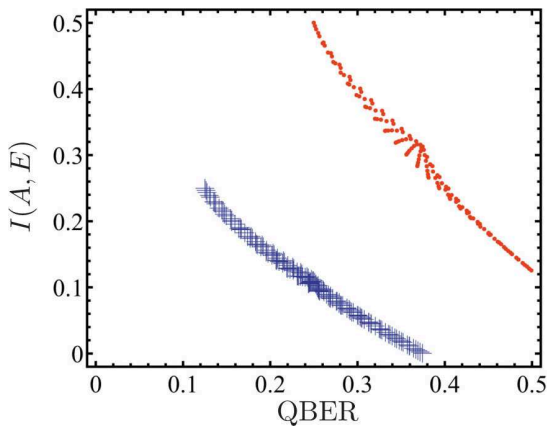}
\caption{\label{csweep} Eve's mutual information on Alice's sifted
key as a function of the induced QBER for different gates $U_2 =
C(\mathbf{c})$. Eve uses the IR attack and measures in the
$\sigma_z$ eigenbasis. The red dots (blue crosses) corresponds to
Eve measuring both (only one) of the two entangled qubits, in
which case Eve's maximal mutual information is between 0.5 and
0.125 (0.25 and 0). }
\end{figure}

Figure~\ref{csweep} shows Eve's mutual information on Alice's
sifted key in the case $N=2$, for an IR attack carried out using the $\sigma_z$
eigenbasis. The plot is obtained by a uniform sweep over the
parameters $\mathbf{c} \in [0,2\pi]^{\times 3}$, over which Alice
can optimize the protocol. In the upper set of points, Eve always
measures both entangled qubits, and in the lower set only one of
them. It makes no difference which qubit is measured, since here
the gate $U_2$ is symmetric with respect to the two entangled
qubits.

The topmost point in each set corresponds to $U_2$ being the
two-qubit identity operation, with which our protocol reduces to
BB84. At the undermost points of the two sets, $U_2 =
C\!\left(0,\frac{\pi}{2},0\right) = U_2^{\displaystyle{\star}} =
\left(I_1\otimes I_1 + i \sigma_y \otimes
\sigma_y\right)/\sqrt{2}$. As $c_2$ increases from 0 to
$\frac{\pi}{2}$, the protocol continuously shifts from BB84 to the
$U_2^{\displaystyle{\star}}$-enhanced protocol.
Eve achieves the maximal information $\frac{1}{2N} = 0.25$ by
changing one of her measurement bases from $\sigma_z$ to
$\sigma_y$.

\begin{figure}[h]
\centering
\includegraphics[width=0.4\textwidth]{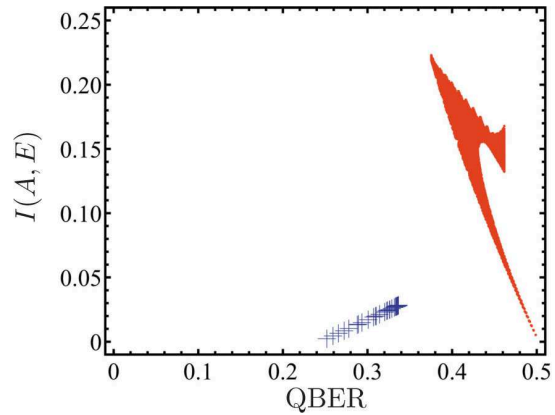}
\caption{\label{evesweep} Eve's mutual information on Alice's
sifted key as a function of the induced QBER sampled over all
possible measurement bases for Eve. The entangling gate is fixed to $U_2 =
C(\mathbf{c}^*)$, where
$\mathbf{c}^*=\left(\frac{\pi}{32},\frac{3\pi}{8},\frac{\pi}{32}\right)$.
The red dots (blue crosses) correspond to Eve measuring
both (only one) of the two entangled qubits, in which case Eve's
maximal mutual information is between 0 and 0.2237 (0 and 0.0284).
}
\end{figure}

Next, we show how to improve on the $1/(2N)$ bound in the case
$N=2$. We allow Eve to use any measurement bases. Thus, the task of
finding the optimal $C(\mathbf{c})$ becomes a twofold optimization
problem: Alice and Bob wish to minimize the maximal information
Eve can obtain for a given QBER. We are thus interested in finding
the value $\min_{\mathbf{c}}
\max_{\{\beta_1,\gamma_1,\beta_2,\gamma_2\}} \left[ I(A,E)/\mathrm{QBER} \right]$
and the optimizing parameter values. We perform the optimization
with the simplex search method \cite{simplex}. One of the optimal
choices of parameters for Alice and Bob is $\mathbf{c}^* =
\left(\frac{\pi}{32},\frac{3\pi}{8},\frac{\pi}{32}\right)$, which
leads to $I(A,E) \approx 0.2237$ and $\mathrm{QBER} = 0.375$ for
$\xi=1$. Given $U_2 = C(\mathbf{c}^*)$, an optimal choice for Eve
is $\left(\beta_1,\gamma_1,\beta_2,\gamma_2\right) =
\left(\frac{\pi}{8},0,\frac{\pi}{2},\frac{\pi}{2}\right)$. Eve's
maximal information as a function of the QBER is shown as the
solid line in Fig.~\ref{infoqber}. For a fixed $\mathrm{QBER} \leq
25\%$, Eve's information drops to less than 30\% of that in BB84.

Figure~\ref{evesweep} elaborates on the consequences of Eve's
choices given $U_2 = C(\mathbf{c}^*)$. In the upper (lower) set of
points, Eve measures both (only one) of the qubits in different
bases. The plot is generated by a uniform sweep over
$\left(\beta_1,\gamma_1,\beta_2,\gamma_2\right) \in
[0,2\pi]^{\times 4}$. Alice's gate is fixed to $C(\mathbf{c}^*)$
which, unlike $U_2^{\displaystyle{\star}}$, is observed not to
guarantee zero but still less than 0.03 bits of information
leakage for one-qubit interceptions.

Let us present an approximate comparison between our protocol and BB84 in terms of the final bit rate. Following Ref.~\cite{bouwmeester}, we assume that during error correction Alice and Bob must exchange
\begin{equation}
nH_{\mathrm{bin}}(q) = n[-q \log_2 q - (1-q)\log_2 (1-q)]
\end{equation}
bits, where $n$ is the length of the key material, and $q$ the QBER. We further make the safe assumption that this is the information, in bits, that is leaked to Eve. In BB84, Eve's information per bit after EC is
\begin{equation}
I_\textrm{EC}^{\mathrm{BB84}}(q) = 2q + H_{\mathrm{bin}}(q).
\end{equation}
Let the optimal $N=2$ setting represent our protocol, where Eve's information after EC is
\begin{equation}
I_\textrm{EC}^{(2)}(q) = s\delta q + H_{\mathrm{bin}}(\delta q),
\end{equation}
where $s = 0.5965$ is the slope of the $I(A,E)$ curve shown in Fig.~\ref{infoqber}. The observed QBER is denoted by $\delta q$, so that $\delta$ is the factor by which the use of $U_2 = C(\mathbf{c}^*)$ changes the QBER. The absolute key rate depends heavily on the practical implementation of the protocol, and we therefore use the relative key rate $r = R_{\mathrm{net}}/R_{\mathrm{sift}}$, where $R_{\mathrm{sift}}$ is the rate at which sifted key bits are generated. We have \cite{gi-rmp-74-145}
\begin{eqnarray}
r(q)    & =& I(A,B) - I(A,E) \nonumber\\
        & =& 1 - H_{\mathrm{bin}}(q) - I(A,E) \nonumber\\
        & =& 1 - I_\textrm{EC}(q)
\end{eqnarray}
for both protocols.

\begin{figure}[h]
\centering
\includegraphics[width=0.4\textwidth]{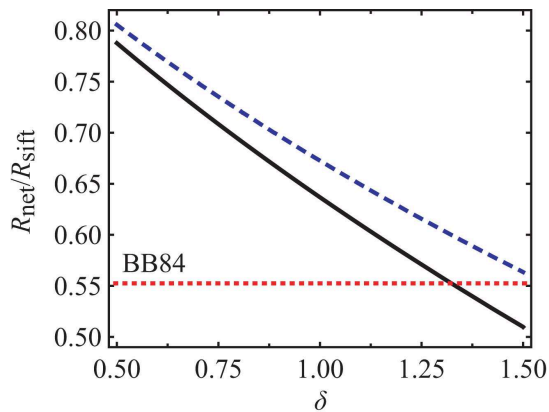}
\caption{\label{rate} Relative key rate as a function of the yet unknown factor $\delta$ by which the scheme changes the QBER. The solid line represents our protocol with $U_2 = C(\mathbf{c}^*)$. The dashed blue line shows the rate for a maximally improving gate, i.e., one which hides all information sent via the quantum channel ($s=0$). The dotted red line shows the rate for BB84. The QBER in BB84 is fixed to 6\%.
}
\end{figure}

In the following, we fix the QBER to $q=6\%$, a typical value in a practical realization \cite{noise1,noise2,noise3,noise4}. Then, the relative key rate is $r_{\mathrm{BB84}} = 0.553$ in BB84. The relative key rate for our two-qubit protocol is shown in Fig.~\ref{rate} together with a protocol, for which $s=0$. For example at $\delta=1$ for both protocols, the gain of the two-qubit protocol over BB84 is 70\% of that of the protocol with $s=0$. The relative key rate of BB84 is recovered at $\delta = 1.323$. Determining the exact value of $\delta$ and ways to decrease it is left for future research.


\section{Conclusions}

Our results show that entanglement can be employed
to considerably improve the BB84-type key
distribution, even in the case of two-qubit entanglement. The new protocol
can be directly adapted to the several variants of BB84. We have
demonstrated one promising scheme, where an IR eavesdropper must
intercept every qubit in the entangled group to gain any information.
Unfortunately, loss of qubits may pose a problem
not only for Eve, but also for Bob. If one of the entangled qubits is completely
lost, the QBER of the remaining qubits is likely to increase.
Therefore, this protocol cannot be recommended for use at extreme
distances where most transmitted qubits are lost \cite{noise4}. Making the
protocol robust against qubit loss is a goal for future research.

Since
the dimension of the total Hilbert space increases exponentially
with the number of qubits, and the dimension of the subspace Eve
can directly access increases only linearly, our scheme is expected
to show even more pronounced benefits if applied to many-qubit
entanglement. Further optimization for an arbitrary number of
entangled qubits and assessment of more potential attacks is to be
carried out in the future. Potential future research also includes
methods for distinguishing between innocent noise in
the quantum channel and that caused
by eavesdropping, and determining the exact dependence of QBER
on the innocent noise. The latter would enable definitive evaluation
of the protocol final bit rate.

\section*{\uppercase{acknowledgments}}
We thank the Academy of Finland, Nokia Corporation, the
Finnish Cultural Foundation, V{\"a}is{\"a}l{\"a} Foundation, the EPSRC, the QIP IRC, and the
Leverhulme Trust for financial support. In addition, Bob Clark is acknowledged for
his warm hospitality at the Centre for Quantum Computer Technology.

\renewcommand{\theequation}{A\arabic{equation}}
\setcounter{equation}{0}
\section*{\uppercase{appendix}}

We show that the gate $U_N^{\displaystyle{\star}}$ defined
in Sec.~\ref{analysis} restricts the information provided by an
intercept-resend attack to at most $1/(2N)$. First, note that
$\sigma_y |a_j;\alpha_j\rangle = |\bar{a}_j;\alpha_j\rangle$
for $j=1,...,N$. We claim that
single-qubit measurements in any basis, applied to state
$U_N^{\displaystyle{\star}}\left(|a_1;\alpha_1\rangle|a_2;\alpha_2\rangle\cdots|a_N;\alpha_N\rangle\right)$,
give
uniformly random results until the last, $N$th, one. Thus it is
not until the last measurement that Eve gets any information
with the IR attack. Let us refer to this randomness of the first
$N-1$ measurements as property $\mathcal{R}$.

To prove this first claim, we note that the transmitted
states for gates $U_{2,\mathrm{even}}^{\displaystyle{\star}}$ and
$U_{2,\mathrm{odd}}^{\displaystyle{\star}}$
are, respectively, $\left(|a_1;\alpha_1\rangle|a_2;\alpha_2\rangle \pm i
|\bar{a}_1;\alpha_1\rangle|\bar{a}_2;\alpha_2\rangle\right)/\sqrt{2}$ and
$\left(|a_1;\alpha_1\rangle|\bar{a}_2;\alpha_2\rangle \pm i
|\bar{a}_1;\alpha_1\rangle|a_2;\alpha_2\rangle\right)/\sqrt{2}$,
on which the first single-qubit measurement, in any basis, gives a
uniformly random result. Hence, gates
$U_{2,\mathrm{even}}^{\displaystyle{\star}}$ and
$U_{2,\mathrm{odd}}^{\displaystyle{\star}}$
have property $\mathcal{R}$.

We write
\begin{equation}
U_{N,\mathcal{P}}^{\displaystyle{\star}} = \frac{1}{\sqrt{2^{N-1}}} \sum_{l=1}^{2^{N-1}} u_{N,\mathcal{P}}^l,
\label{gatesum}
\end{equation}
where the parity $\mathcal{P}$ is even or odd, and
$u_{N,\mathcal{P}}^l$ is a unique tensor product of operators
$I_1$ and $\sigma_y$, with an even or odd number of operators
$\sigma_y$ according to the parity $\mathcal{P}$. The parity is
invariant under the application of Eq.~(\ref{ngate}). As the total
number of different $N$-qubit tensor products of $I_1$ and
$\sigma_y$ is $2^N$ and half of them have an even number of
operators $\sigma_y$, the sum in Eq.~(\ref{gatesum}) contains all
possible $u_{N,\mathcal{P}}^l$ of the given parity $\mathcal{P}$.
It follows that any permutation of qubits in the state
$U_{N,\mathcal{P}}^{\displaystyle{\star}}
\left(|a_1;\alpha_1\rangle \cdots |a_N;\alpha_N\rangle\right)$
results in essentially the same state, i.e., only the phases of
the different terms change, which has not effect on the outcome of
the following measurement. Hence, we can assume that the leftmost
qubit is measured first, without restricting Eve's actual order of
measurements. Thus the application of the gate
$U_N^{\displaystyle{\star}}$ is not limited to IR attack, for
which the measurement order of the eavesdropper is determined by
Alice.

According to Eq.~(\ref{ngate}), the outcome of measuring the leftmost qubit in the state $U_{N,\mathcal{P}}^{\displaystyle{\star}} \left(|a_1;\alpha_1\rangle \cdots |a_N;\alpha_N\rangle\right)$ is uniformly random. Moreover, a correct result leads to the remaining state to be that resulting from application of gate $U_{N-1,\mathcal{P}}^{\displaystyle{\star}}$, i.e., the gate of the same parity. An incorrect result leads to the state corresponding to $U_{N-1}^{\displaystyle{\star}}$ of different parity. Thus, gate $U_N^{\displaystyle{\star}}$ has property $\mathcal{R}$ for all $N>1$. As Eve measures the qubits, she unwinds the recursion of Eq.~(\ref{ngate}) through even and odd states while learning nothing of the key until the remaining state has $N=1$.

Let $E_1$ and $E_N$ denote the random variables of the outcomes of
the first $N-1$ measurements and the final, $N$th measurement Eve
makes, respectively. Denote the conditional entropy of $E_N$ as
$h_N = H(E_N|A,E_1)$. Note that $0 \leq h_N \leq 1$. If
$U_N^{\displaystyle{\star}}$ is used, $I(A,E_1)=0$. The entropy
$H(E_1) = N-1$. Using the definition of conditional entropy
$H(X|Y) = H(X,Y) - H(Y)$, we obtain
\begin{eqnarray}
I(A,E)  &=& \frac{1}{N}\left[2N - H(A,E_1,E_N)\right] \nonumber \\
            &=& \frac{1}{N}\left[2N - h_N - H(E_1|A) - H(A)\right] \nonumber \\
            &=& \frac{1}{N}\left[N - h_N - H(E_1)\right] \nonumber \\
            &=& \frac{1}{N}\left(1 - h_N\right),
\end{eqnarray}
where we have recomposed random variables as $I(A,E) =
I(A,E_1,E_N) = I(AE_1,E_N)$. Since the last measurement targets
one qubit in a BB84 state, $h_N = \frac{1}{2}$ and $I(A,E) =
\frac{1}{2N}$. This completes our proof that the gate
$U_N^{\displaystyle{\star}}$ limits the information provided by an
intercept-resend attack to at most $1/(2N)$ per bit.

\bibliography{bib17}

\end{document}